\documentclass[DEEPLIST,oneside,12pt,doublespacing]{elsart}
\usepackage{natbib}
\usepackage{graphicx}
\begin{document}
\begin{frontmatter}
\title{Soft X-ray emissions of Si~IX in Procyon}
%   \volnopage{Vol.0 (200x) No.0, 000--000}      %%preserved for Editor. DOn't remove!
%   \setcounter{page}{1}          %%starting page, preserved for Editor. DOn't remove!
\author[Liang]{Guiyun Liang} ~~and~~
\author[Liang]{Gang Zhao}~$^{\ast}$
\corauth[Liang]{Corresponding author: Gang Zhao}
\ead{gzhao@bao.ac.cn}
\address[Liang]{National Astronomical Observatories, Chinese
Academy of Sciences, Beijing
 100012, P. R. China}

\begin{abstract}
\baselineskip=18pt An analysis of $n=3 \to 2$ transition lines of
carbon-like silicon reveals that some ratios of line intensities
are sensitive to the electron density. The ratio between two group
of $3d\to2p$ transition lines at 55.246~\AA\, and 55.346~\AA\, is
a good $n_{\rm e}$-diagnostic technique, due to its insensitivity
to the electron temperature. Using this property, a lower limit of
the density of 0.6$\times10^8$cm$^{-3}$ is derived for Procyon,
which is consistent with that constrained by C V and Si X
emissions. Significant discrepancies in ratios of $3s\to2p$ lines
to $3d\to2p$ lines between theoretical predictions and observed
values, are found, by the spectral analysis of Procyon observed
with the {\it Chandra} High Resolution Transmission Grating
spectra. The difference exceeding a factor of 3, cannot be
explained by the uncertainty of atomic data. The opacity effect is
also not a choice as reported by Ness and co-workers. For the
$3s\to2p$ line at 61.611~\AA\,, present work indicates that the
large discrepancy may be due to the contamination from a S VIII
line at 61.645~\AA\,. For the lines at 61.702 and 61.846~\AA\,, we
suggest that the discrepancies may be attributed to contaminations
of unknown lines.
\newline \vspace{2cm}
\end{abstract}
\end{frontmatter} \maketitle

\section{Introduction}
Since the launch of the new generation X-ray observatories, {\it
Chandra} and {\it XMM-Newton}, X-ray spectra with unprecedented
spectral resolution and high statistical quality can be obtained
for astrophysical objects. The High Energy Transmission Grating
Spectrometer (HETGS) and Low Energy Transmission Grating
Spectrometer (LETGS) on {\it Chandra} can provide resolving power
of $\lambda/\Delta\lambda~\sim~100-1000$ and $\ge1000$ in
wavelength ranges of 1.5-30\AA\, and 50-160\AA, respectively. The
effective areas can be as large as 200~cm$^2$ in some energy
regions. Using these high-quality spectra, physical conditions of
stellar coronae such as electron density, thermal structure, and
element abundances, can be derived. Such information further helps
to constrain models of the coronal heating and structuring. In the
past, the information about the spatial distribution of coronal
plasma was inferred by indirect means such as modelling of
eclipses and rotational modulation (G$\ddot{\rm u}$del et al.,
1995, 2003; Siarkowski et al., 1996), and such analyses can only
be carried out for very special systems with advantageous
geometries. Presently, the electron density ($n_{\rm e}$) of
stellar coronae can be disentangled from the emission measure
($EM=\int n_{\rm e}^2dV$), and determined separately using the
X-ray spectra with high-resolution for various late-type stars.

The complexes of He-like ``triplets" of C, N, O, Ne, Me, and Si,
which include the {\it resonance} ($r$: $1s^2~^1S_0 -
1s2p~^1P_1$), {\it inter-combination} ($i$: $1s^2~^1S_0 -
1s2p~^3P_{2,1}$), and {\it forbidden} ($f$: $1s^2~^1S_0 -
1s2s~^3S_1$) lines, are covered by the HETGS and/or LETGS on {\it
Chandra} and Reflection Grating Spectrometer (RGS) on {\it
XMM-Newton}, and are resolvable. Gabriel \& Jordan (1969) pointed
out the potential diagnostic application of these features for hot
plasmas. Recently, Porquet \& Dubau (2000) recalculated line
intensities by accounting for other processes such as dielectronic
recombination and radiative recombination. In the HETGS
observation of Capella, Canizares et al. (2000) estimated a
density range of 0.8-2$\times10^{10}$~cm$^{-3}$ from O~VII, and
upper limits near 7$\times10^{11}$ and 1$\times10^{12}$~cm$^{-3}$
by Mg~XI and Si~XIII. Using the LETGS observation of Capella,
Brinkman et al. (2000) derived a density of
$\sim2.6\times10^9$~cm$^{-3}$ by C V for the lower temperature
component of a multi-temperature structure. Later, Ness et al.
(2002) and Testa et al. (2004) performed coronal density
diagnostics with He-like triplets for other late-type stars with
various activities such as Algol, Procyon, $\alpha$~Cen~A\&B,
$\epsilon$~Eri, and HR~1099 etc. Using the derived densities, Ness
et al. (2004) further estimated the sizes of stellar X-ray
coronae, and concluded that the cooler plasma component cannot
cover a large fraction of the stellar surface, and there must be
spatially separate plasma components at different temperatures.
They also argued that the hotter plasma loops fill the space
between cooler loops until much of the corona is dominated by the
hotter plasma.

The parameter of the electron density is also used to judge the
mechanism of the X-ray production. For T~Tauri stars, a density of
$>1\times10^{12}$~cm$^{-3}$ has been obtained from O~VII for
TW~Hya (Stelzer \& Schmitt, 2004) and BP~Tau (Schmitt et al.,
2005), which is higher than typical coronal densities by at least
two orders of magnitude. It was assumed that the X-ray emission is
produced in the accretion shocks of T~Tauri stars, instead of
stellar coronae. Additionally, the UV radiative field from the
inner layer also plays an important role on the population of
levels $^3P_{2,1}$, which directly influences the estimation of
$n_{\rm e}$. For lighter elements, this effect is more sever (Ness
et al., 2001). In the coronal X-ray spectra, in addition of the
emissions of H-, He-like ions and Fe L-shell ions, many lines from
highly charged silicon have been observed in the LETGS
observations, such as that of Procyon. These emissions may give
further insight into the physical conditions of stellar coronae.
Recently, we estimated the electron densities and/or temperatures
for the cooler components using Si~X and Si~XI features (Liang et
al. 2006a). The emissions from $3d\to2p$ and $3s\to2p$ transitions
of Si~IX have also been identified by Raassen et al. (2002) in the
LETGS observation of Procyon. In this paper, we investigate the
soft X-ray emissions of Si~IX in detail.

This paper is organized as followings, observations and
measurements of line fluxes are described in Sect.2. Theory model
of C-like Si~IX is presented in Sect.3. Sect.4 outline the results
and discussions, and the conclusion is given in the last section.

\section{Observations and line flux measurements}
Procyon (F5~IV--V) is a solar-like star at a distance of 3.5pc,
with a mass of 1.7$M_{\odot}$ and a radius of 2.06$R_{\odot}$. So
far, three different observations with Low Energy Transmission
Grating Spectrometer (LETGS) combined with High Resolution Camera
(HRC) are available for this star in the {\it Chandra} Public Data
Archive.\footnote{http://cxc.harvard.edu/cda/} The observations
are performed on three days, which are summarized in Table 1. We
utilized the CIAO3.3
software\footnote{http://cxc.harvard.edu/ciao/download/} and some
science threads for the LETGS/HRC observation to reduce the
observation data. In order to increase the signal-to-noise ratio
(SNR), we co-added these three different observations, resulting
in a spectrum with total exposure time of 159.9ks after times of
bad counts were excluded.
\begin{table}[]
  \caption[]{The properties of three different observations of Procyon with
LETGS/HRC.}
    \begin{center}\begin{tabular}{cccc}
    \hline\noalign{\smallskip}
{\rm Sequence} & {\rm Obs\_ID} & {\rm Exposure~time} & {\rm Start~time}\\
{\rm number} &  & {\rm (ks)} &  \\
      \hline\noalign{\smallskip}
290032 & 63 & 70.15 & 1999.11.6~~21:10:29 \\
280411 & 1461 & 70.25 & 1999.11.7~~16:58:44  \\
280174 & 1224 & 20.93 & 1999.11.8~~12:38:44  \\
  \noalign{\smallskip}\hline
  \end{tabular}\end{center}
\end{table}

Spectral line fluxes were measured using the {\it Sherpa} software
package in the CIAO3.3. Gaussian profile was adopted here, because
it can precisely describe the line broadening due to instruments
for point-like sources. In the measurements of line fluxes and the
determinations of element abundances, the continuum emission plays
an important role. It is very difficult to accurately determine
the continuum emission in active stars due to its relatively high
level. Normal stars, such as Procyon, show very low continuum
emission, and do not introduce large uncertainties in the measured
line ratios. Here, a constant value of
1.34$\times10^{-4}$phot.s$^{-1}$cm$^{-2}$\AA$^{-1}$ was adopted to
represent the source continuum level, which was determined from
the line free region, such as in 51--52\AA\,.
%%%%%%%%%%Figure-1%%%%%%%%%%%%%%%%%%%%
\begin{figure}
\vspace{2mm}
   \begin{center}
   \hspace{3mm}\includegraphics[width=10cm,height=7.5cm]{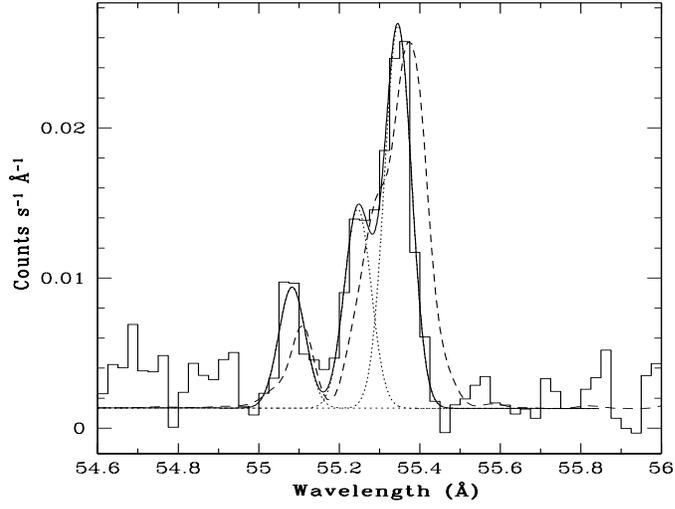}
   \parbox{180mm}{{\vspace{2mm} }}
   \caption{Observed spectrum of Procyon in wavelength range of 54.6--55.0\AA\,, and
best-fit result (solid smooth line) from fitting with three
Gaussian profiles (dotted lines) and a constant value. Dashed line
represents the synthesized spectrum.}
   \end{center}
\end{figure}

%%%%%%%%%%Figure-2%%%%%%%%%%%%%%%%%%%%
\begin{figure}
\vspace{2mm}
   \begin{center}
   \hspace{3mm}\includegraphics[width=10cm,height=7.5cm]{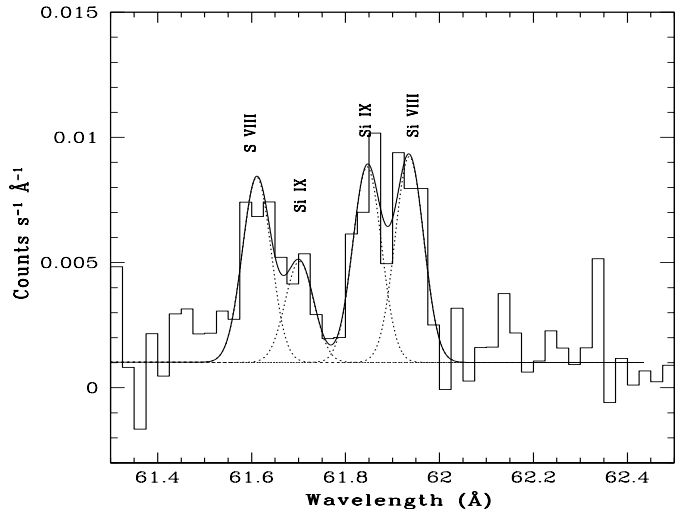}
   \parbox{180mm}{{\vspace{2mm} }}
   \caption{Observed spectrum of Procyon in wavelength range of 61.5--62.5\AA\,, and
best-fit result (solid smooth line) from fitting with four
Gaussian profiles (dotted lines) and a constant value.}
   \end{center}
\end{figure}

Figs.1 and 2 illustrate the observed spectra with background
subtracted in ranges of 54.6--56.0 and 61.5--62.5\AA\, regions,
respectively. For lines around 55.0\AA\,, only the positive order
spectrum was adopted because of a gap around 54\AA\, in negative
order spectrum, while for lines around 62.0\AA\,, the negative
order spectrum was used due to the gap around 61\AA\, in the
positive order spectrum. Best-fit results (smooth solid lines) are
overlayed in the two figures. In the fitting, three and four
components with same FWHM were used in the two wavelength ranges,
respectively. The line centroid, amplitude of each component, and
FWHM are free parameters. Visual inspection and reduced $\chi^2$
(1.1 and 1.3) indicate that the best-fit results are acceptable,
which are overlayed in Figs.1--2. Although the resulted line
fluxes are slightly over-estimated, they are still within the
range of statistical uncertainties~(1$\sigma$). The line fluxes
and identification for features around 55.0 and 62.0\AA\, are
listed in Table 2.

\begin{table}
  \caption[]{Observed wavelength and flux (in unit of $\times10^{-4}$phot. cm$^{-2}$s$^{-1}$)
    of prominent features in ranges of 54.6--56.0\AA\, and 61.5--62.5\AA\,, and their identification.}
  \begin{center}\begin{tabular}{ccccccllll}
    \hline\noalign{\smallskip}
{\rm Index} & ${\rm \lambda_{obs}~(\AA)}$ & {\rm Ions}& ${\rm \lambda_{theo}~(\AA)}$ & ${\rm Flux_{obs}}$ & ${\rm Flux_{theo}}$  & \multicolumn{4}{c}{\rm Transition} \\
     \hline\noalign{\smallskip}
1a & 55.083 & {\rm Si~IX} & 55.094 & 0.68$\pm$0.12 & 0.11 & $2s^22p3d$ & $^3P_0$ &$\to~~2s^22p^2$ & $^3P_1$ \\
1b & ...    & {\rm Si~IX} & 55.116 & ...         & 0.14 & $2s^22p3d$ & $^3P_1$ &$\to~~2s^22p^2$ & $^3P_1$ \\
2a & 55.246 & {\rm Si~IX} & 55.234 & 1.12$\pm$0.14 & 0.12 & $2s^22p3d$ & $^3P_1$ &$\to~~2s^22p^2$ & $^3P_2$ \\
2b & ...    & {\rm Si~IX} & 55.272 & ...         & 0.34 & $2s^22p3d$ & $^3P_2$ &$\to~~2s^22p^2$ & $^3P_2$ \\
2c & ...    & {\rm Si~IX} & 55.305 & ...         & 0.41 &$2s^22p3d$ & $^3D_1$ &$\to~~2s^22p^2$ & $^3P_0$ \\
3a & 55.346 & {\rm Si~IX} & 55.356 & 2.16$\pm$0.17 & 0.76 & $2s^22p3d$ & $^3D_2$ &$\to~~2s^22p^2$ & $^3P_1$ \\
3b & ...    & {\rm Si~IX} & 55.383 & ...         & 0.15 & $2s^22p3d$ & $^3D_1$ &$\to~~2s^22p^2$ & $^3P_1$ \\
3c & ...    & {\rm Si~IX} & 55.401 & ...         & 0.70 & $2s^22p3d$ & $^3D_2$ &$\to~~2s^22p^2$ & $^3P_2$ \\
4a & 61.611 & {\rm Si~IX} & 61.600 & 0.75$\pm$0.15 & 0.04 & $2s^22p3s$ & $^3P_1$ &$\to~~2s^22p^2$ & $^3P_0$ \\
4b & ...    & {\rm S~VIII}& 61.645 & ...         & & $2s^22p^43s$ & $^2D_{5/2}$ & $\to ~~ 2s^22p^5$ & $^2P_{3/2}$\\
4c & ...    & {\rm Si~IX} & 61.649 & ...         & 0.12 & $2s^22p3s$ & $^3P_2$ &$\to~~2s^22p^2$ & $^3P_2$ \\
5  & 61.702 & {\rm Si~IX} & 61.696 & 0.41$\pm$0.14 & 0.03& $2s^22p3s$ & $^3P_1$ & $\to ~~2s^22p^2$ & $^3P_1$\\
6  & 61.846 & {\rm Si~IX} & 61.844 & 0.79$\pm$0.16 & 0.05& $2s^22p3s$ & $^3P_1$ &$\to~~2s^22p^2$ & $^3P_1$ \\
7  & 61.936 &{\rm Si~VIII}& 61.914 & 0.83$\pm$0.16 & & $2s^22p^2(^1D)3d$ & $^2F_{7/2}$ &$\to~~2s^22p^3$ & $^2D_{5/2}$ \\
  \noalign{\smallskip}\hline
  \end{tabular}\end{center}
  \flushleft{{\it Notes:} The same labels with different lowercase
indices (e.g., 3a, 3b and 3c) indicate blended lines.}
\end{table}

\section{Atomic model of Si~IX}
In our model, the excitations and de-excitations induced by
energetic electrons and the subsequent radiative decays among 484
levels belonging to 28 configurations [namely, ($1s$)$2s^22p^2$,
$2s2p^3$, $2p^4$, $2s^22p3l$, $2s2p^23l$, $2s^22p4l$, $2s2p^24l$,
$2s^22p5l$ and $2s^22p6l$ ($l=0, 1, ..., n-1$)] have been
considered. The optical thin approximation combined with the
collision ionization equilibrium can perfectly describe the
coronal plasma, and are used extensively in astrophysical
modelling, and we adopt this assumption as well. The atomic data,
which was calculated using the FAC package~(Gu 2003), has been
assessed in our previous paper~(Liang et al. 2006b). For lower
levels, the level populations induced by the collision between the
proton and the highly charged Si~IX ion, are non-negligible, and
we adopt the rates of Rayans et al. (1999) in present analysis.
For some weaker transitions, the resonant excitation (RE) process
contributes tens of percents to the population. In principle, the
excitation rates including resonant effect for all transitions
should be included. However, for the strong lines we are
interested in, the effects of RE appears to be small, and such a
model is not necessary. Moreover, the calculation of excitation
rates including RE is very difficulty for large-scale atomic
model. So far, only a few excitation data of Si~IX including RE
are available, and they are for transitions among lowest levels
with $n=2$ configurations, which ware used in our model.

For ions with an open valence shell, such as Si~IX, competing
between population and de-population for each energy level cause
the populations of some levels to be sensitive to the electron
density, $n_{\rm e}$. Accordingly, some line intensity ratios may
also be sensitive. In case of Si~IX, we found the ratio between
two group of $3d\to2p$ transition lines, the 55.246\AA\, to
55.346\AA\, (the values refer to the resolvable observed
wavelengths in Table 2), is sensitive to $n_{\rm e}$, as shown in
Fig.3. The blending effect from all significant features (refer to
Table 2) has been considered in the prediction of the ratio
(hereafter $R$). There are two advantages for $n_{\rm
e}$-diagnostic using this ratio. One is that the two resolved
emissions are the strongest emissions, so they have relatively
higher SNR, and the ratio is around unity. The other
characteristic is that the ratio is insensitive to the electron
temperature $T_{\rm e}$ as shown in Fig.3. The variation of the
ratio is less than 3\% when the electron temperature changes
0.2dex. For coronal plasmas (log$n_{\rm e}\geq7.5$), the
temperature sensitivity is even weaker.

%%%%%%%%%%Figure-3%%%%%%%%%%%%%%%%%%%%
\begin{figure}[h]
\vspace{2mm}
   \begin{center}
   \hspace{3mm}\includegraphics[angle=-90,width=9.5cm]{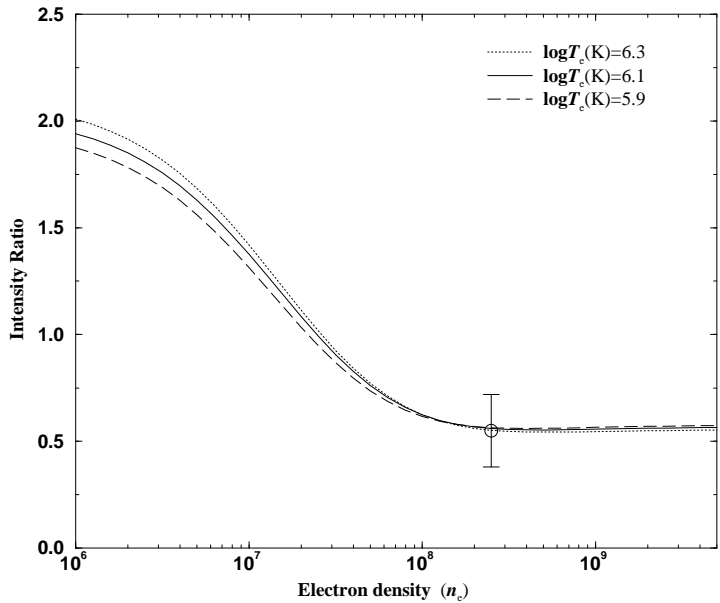}
   \parbox{180mm}{{\vspace{2mm} }}
   \caption{Predicted line ratio $R$ of two resolvable lines at
55.246 and 55.346\AA\,, as a function of the electron density
$n_{\rm e}$ at three different logarithmic temperatures (in K) of
5.9 (dash), 6.1 (solid) and 6.3 (dot). The symbol with error bar
refers its observed value in Procyon.}
    \end{center}
\end{figure}

\section{Results and discussions}
\subsection{Line identification}
In the selected wavelength range, forest-like emissions from
highly charged silicon, magnesium, sulfur, calcium and iron are
present. Although we disentangled those partially blended
emissions through fitting with multi-components, there are several
individual lines in each isolated feature. As listed in Table 2,
almost all features in this wavelength region, are contaminated.
For the line at 55.083\AA\,, a feature of Mg~IX (55.060\AA\,) may
be a source of contribution as reported by Raassen et al. (2002).
For the lines at 55.246 and 55.346\AA\,, several $3d\to2p$
transitions of Si~IX are identified. Fortunately, no emissions
from other charged stages and elements contaminate them. Analysis
based on the line ratio technique for the two groups avoids
uncertainties of the element abundance and ionization equilibrium.
The blending effect due to $3d\to2p$ transitions can be considered
in a single-ion model. For the line at 61.611\AA\,, Raassen et al.
(2002) measured a wavelength of 61.578\AA\, and assigned it to a
S~VIII line with 61.600\AA\,. The difference in the measured
values of the wavelength might be from the statistic uncertainties
and the use of updated calibration files in the present analysis.
In the recent experimental work of Lepson et al. (2005), a
$3s\to2p$ transition line of S~VIII is detected at the wavelength
of 61.645\AA\, with a strength of 0.75 relative to that of
$3d\to2p$ transition line at 52.781\AA\,, and it is the strongest
in the local wavelength range. The obvious detection of $3d\to2p$
line in Procyon, indicates that a fraction of the observed flux
around 61.611\AA\, may be attributed to the S~VIII line. A search
in APEC1.3.1 model (Smith 2001), shows that the emissivity of
Si~IX line at 61.649\AA\, is the strongest in this wavelength
range. Present model shows a similar result. We also note that the
emissivity of Si~IX line at 61.696\AA\, is $\sim$0.18 relative to
that of 61.649\AA\, line, so we attribute it to be the possible
emission at 61.702\AA\,, whereas Raassen et al. (2002) measured a
wavelength of 61.668\AA\, and assigned to be a different feature
of Si~IX at 61.649\AA\,. The feature at 61.846\AA\, is presently
assigned to be $3s\to2p$ transition of Si~IX (61.844\AA\,),
because no other prominent transitions are close to this
wavelength. The observed line at 61.936\AA\, is assigned to
Si~VIII as reported by Raassen et al. (2002).

3-$T_{\rm e}$ CIE model can describe the observed spectrum
satisfactorily as performed by Raassen et al. (2002). One
component (with $T_{\rm e}$=1.21$\pm0.07$~MK and
$EM=2.45\pm0.27\times10^{50}$~cm$^{-3}$) of the multi-temperature
model is very close to the peak temperature (1.26~MK) of maximum
Si~IX fraction in the ionization equilibrium (Mazzotta et al.
1998). In our previous work (Liang et al. 2006a), we derived a
density of 2.6$\times10^8$~cm$^{-3}$ for the cooler plasma of
Procyon, using lines of Si~X. Using the electron temperature and
density, as well as EM and the element abundance estimated by
Raassen et al. (2002) for LETGS observation of Procyon, we
calculate the theoretical line fluxes as listed in Table 2. The
differences between the prediction and observation of line fluxes
are primarily from the uncertainty of the element abundance,
because their result is from the global fitting of the Procyon
spectrum. For $3d\to 2p$ transitions, good agreements are
obtained.

Using Gaussian line profile, we construct a synthesized spectrum
of Si~IX. The theoretical spectrum normalized to observed line at
55.346\AA\, is overlayed on the observed spectrum, as shown by the
dashed line in Fig.1. The comparison reveals that the three
identified wavelengths around 55.346\AA\, appear to have slight
more separation in wavelengths, which result in the theoretical
spectrum appears broader at this position. This may be from the
uncertainty of the atomic data, although the upper levels for
these transitions are from
NIST\footnote{http://physics.nist.gov/cgi-bin/AtData/main\_asd}.
For line at 55.083\AA\,, Raassen et al. (2002) assigned the line
flux to be partially from Mg~IX emission at 55.050\AA\,, while its
contribution is very weak as revealed by the present model.

\subsection{Estimation of $n_{\rm e}$}
From the subtracted line fluxes in Procyon, we obtain the observed
value (5.2$\pm$0.11) for the ratio $R$. The comparison of the
predicted $R$ with the observed value indicates that the electron
density is about $>0.6\times10^8$cm$^{-3}$ in the line emitting
region.

In collision ionization equilibrium (Mazzotta et al. 1998), the
peak temperature (1.26MK) of Si~IX fraction is close to the
$T_{\rm e}$ determined from H- and He-like carbon ions. Therefore,
the densities determined by the two different methods should be
comparable. However, the radiation effect from Procyon's
photosphere has great influence on the estimation of $n_{\rm e}$
using the triplet ratio of He-like carbon. When this effect has
been considered, an upper limit of $< 8.3\times10^8$cm$^{-3}$ was
obtained (Ness et al. 2002). The combination of the two results
definitely gives the density range for the cooler X-ray emitting
region. This ratio can also be used to diagnose the $n_{\rm e}$
for those C-depleted stars, such as Algol.

\subsection{Line ratios of $3s\to2p$ {\it vs} $3d\to2p$}
Ratios of emission with small oscillator strength $f$ {\it vs}
emission with large oscillator strength have been used to detect
resonant scattering. For example, the ratio of the $3s\to2p$ line
of Fe~XVII at 16.78\AA\, ($gf$=0.01) to $3d\to2p$ line at
15.03\AA\, ($gf$=2.66) has been used to infer the effects of
resonant scattering, because even the most complete model of Doron
\& Behar (2002) do not agree with the observations, such as the
flare observation of AB~Doradus (Matranga et al. 2005). Large
discrepancies in the predicted and observed ratios for similar
lines in Fe~XVIII and Fe~XIX have also been reported by Desai et
al. (2005).

In Procyon, two resolved $3s\to2p$ lines have been identified in
the work of Raassen et al. (2002). However, the identification is
not conclusive due to the uncertainty of atomic data. Here we make
a detailed analyses for these possible $3s\to2p$ emissions. For
the emission at 61.611\AA\,, the previous subsection has revealed
that a fraction of the observed flux originates from the emission
of S~VIII line at 61.645\AA\,. In order to subtract the
contribution of S~VIII emission line, we adopt the experimental
line intensity ratio (0.75) between 61.645 and 52.781\AA\,,
measured in EBIT facility by Lepson et al. (2005). The observed
flux of the obvious emission around 52.780\AA\, is
0.39$\times10^{-4}$phot.cm$^{-2}$s$^{-1}$. So S~VIII occupies
$\sim$39\% around 61.611\AA\,.  By taking into account this
blending, the observed intensity at 61.611\AA\, is 0.21$\pm0.07$
relative to that of the line 55.346\AA\, as shown by the circle
symbol in Fig.~4, which is slightly higher than the prediction
over a large range of density. However, they agree within
2$\sigma$ statistical errors. For the other two emissions at
61.702 and 61.846\AA\,, only the features of Si~IX from $3s\to2p$
transitions at 61.696 and 61.844\AA\, can be found in the present
prediction and available database such as APEC and MEKAL. When the
line intensities are normalized to that of 55.346\AA\, line, the
observed ratios are significantly higher than the theoretical
prediction (solid line) by an order of magnitude in typical
coronal conditions, as shown by square and diamond symbols with
$1\sigma$ error in Fig.~4.
%%%%%%%%%%Figure-4%%%%%%%%%%%%%%%%%%%%
\begin{figure}[h]
\vspace{2mm}
   \begin{center}
   \hspace{3mm}\includegraphics[angle=-90,width=9.5cm]{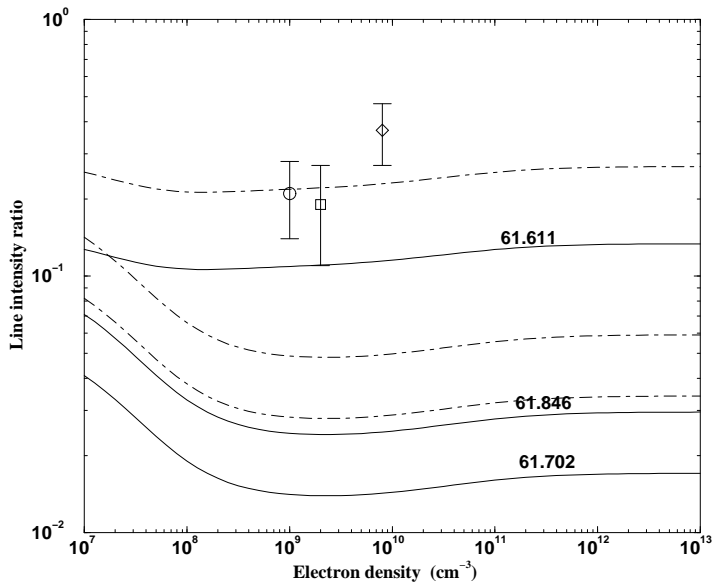}
   \parbox{180mm}{{\vspace{2mm} }}
   \caption{Observed line ratios with $1\sigma$ error (symbols) of identified $3s\to2p$
lines relative to that of the line at 55.346\AA\,. Theoretical
predictions at the temperature log$T_{\rm e}$(K)=6.1 are
overlayed. The dot-dash lines represent agreement within a factor
of 2. }
    \end{center}
\end{figure}

A search in the Astrophysical Emission Code (APEC with version of
1.3.1) also indicates the emissivities of Si~IX lines are the
strongest. So the differences must be attributed to other
explanations. The first thought is that whether the discrepancies
are due to the contaminations from high-order spectral lines. In
the spectrum of Procyon, no second- and third-order spectral lines
of prominent lines in short wavelength range ($<$30\AA\,) lie in
this wavelength region. This encourages us to consider the
incompleteness of model and/or the opacity effect. In the case of
Fe~XVII, Doron \& Behar (2002) considered the effects of
recombination, resonance excitation and ionization processes on
the level populations, and made the theoretical prediction
increase by 50\%. The correct results shown a better agreement
with observed ones for most stellar coronae.  In the case of
Si~IX, observed ratio at 61.611\AA\, can agree with the prediction
within $1\sigma$ uncertainty, if the indirect processes have a
similar effect as in Fe~XVII. For the other two possible $3s\to2p$
emissions, the discrepancies being up to an order of magnitude,
cannot be explained by the complete model. As suggested by Ness et
al. (2003), the opacity effect is also not possible to be detected
from the integrated spectrum for normal stars. So we suggest that
the most possible reason might be the contamination of unknown
emissions.

\section{Conclusions}
A detailed analysis of soft X-ray emissions of carbon-like silicon
reveals that the ratio of Si~IX lines around 55.246 and
55.346~\AA\, is a good $n_{\rm e}$-diagnostic method. By comparing
the observed ratio with theoretical prediction, a lower limit of
the density (0.6$\times10^8$~cm$^{-3}$) was obtained for Procyon,
which is the typical value of the solar quiescent corona. The
constructed spectrum of Si~IX suggests that the major contribution
at 55.083\AA\,, and 61.611\AA\, should be assigned to Si~IX.

In intensity ratios of $3s\to2p$ transition lines of Si~IX
relative to that of $3d\to2p$ line at 55.346~\AA\,, significant
discrepancies between observations and theoretical predictions are
found in Procyon. For the $3s\to2p$ line at 61.611~\AA\,, the
observation and theoretical prediction agree within a factor of 2,
when the contamination arising from S~VIII line (61.645~\AA\,) has
been considered. Here, an experimental ratio of 0.75 (Lepson et
al. 2005) between S~VIII lines at 61.645 and 52.781~\AA\, was
adopted to disentangle the blending. However, for other two
possible $3s\to2p$ lines of 61.792 and 61.848~\AA\,, large
differences (nearly an order of magnitude) cannot be explained by
the uncertainty of atomic data and the resonant scatter effect. So
we suggest that the most possible reason is contributions of
unknown emissions, which may be addressed by further improvements
in the soft X-ray spectral analysis.

\section*{Acknowledgments}
This work is supported by the National Natural Science Foundation
of China under Grant No. 10433010, 10403007 and 10521001.


\begin{thebibliography}{99}
\bibitem{BGK00}
      Brinkman, A. C., Gunsing, C. J. T., Kaastra, J. S., et al.
      2000, ApJ, 530, L111
\bibitem{CHD00}
      Canizares, C. R., Huenemoerder, D. P., Davis, D. S., et al.
      2000, ApJ, 539, L41
\bibitem{DBD05}
      Desai, P., Brickhouse, N. S., Drake, J. J., et al. 2005,
      ApJ, 625, L59
\bibitem{DB02}
      Doron, R., \& Behar, E. 2002, ApJ, 574, 518
\bibitem{GJ69}
      Gabriel, A. H., \& Jordan, C. 1969, MNRAS, 145, 241
\bibitem{Gu03}
      Gu, M. F. 2003, ApJ, 582, 1241
\bibitem[G$\ddot{\rm u}$del et al., 2003]{gued03}
      G$\ddot{\rm u}$del, M., Arzner, K., Audard, M., \& Mewe, R.
      2003, A\&A, 403, 155
\bibitem[G$\ddot{\rm u}$del et al., 1995]{gued95}
      G$\ddot{\rm u}$del, M., Schmitt, J. H. M. M., Benz, A. O.,
      \& Elias, N. M. 1995, A\&A, 301, 201
\bibitem{LBB05}
      Lepson, J. K., Beiersdorfer, P., Behar, E., \& Kahn, S. M.
      2005, ApJ, 625, 1045
\bibitem{LZS06a}
      Liang, G. Y., Zhao, G., \& Shi, J. R. 2006a, AJ, 132, 371
\bibitem{LZZ06}
      Liang, G. Y., Zhao, G., \& Zeng, J. L. 2006b, Atom. Data and
      Nucl. Data Tables (accepted)
\bibitem{MH}
      Matranga, M., Mathioudakis, M., Kay, H. R. M., \& Keenan, F.
      P. 2005, ApJ, 621, L125
\bibitem{MMC98}
     Mazzotta, P., Mazzitelli, G., Colafrancesco, S., \& Vittorio,
     N. 1998, A\&AS, 133, 403
\bibitem{NGS04}
      Ness, J. -U., G$\ddot{\rm u}$del, M., Schmitt, J. H. M. M., Audard, M., \& Telleschi, A. 2004,
      A\&A, 427, 667
\bibitem{NMS01}
      Ness, J. -U., Mewe, R., Schmitt, J. H. M. M. et al. 2001,
      A\&A, 367, 282
\bibitem{NSA03}
      Ness, J. -U., Schmitt, J. H. M. M., Audard, M., G$\ddot{\rm
      u}$del, M., \& Mewe, R. 2003, A\&A, 407, 347
\bibitem{NSB02}
      Ness, J. -U., Schmitt, J. H. M. M., Burwitz, V., et al.
      2002, A\&A, 394, 911
\bibitem{PD00}
      Porquet, D., \& Dubau, J. 2000, A\&AS, 143, 495
\bibitem{RMA02}
      Raassen, A. J. J., Mewe, R., Audard, M., et al. 2002, A\&A,
      389, 228
\bibitem{RFK99}
      Ryans, R. S., Foster-Woods, V. J., Keenan, F. P., \& Reid, R.
      H. G. 1999, Atom. Data and
      Nucl. Data Tables, 73, 1
\bibitem{SRN05}
      Schmitt, J. H. M. M., Robrade, J., Ness, J. -U., et al.
      2005, A\&A, 432, L35
\bibitem[Siarkowski et al., 1996]{SPD96}
      Siarkowski, M., Pres, P., Drake, S. A., White, N. E., \&
      Singh, K. P. 1996, ApJ, 473, 470
\bibitem{SS04}
      Stelzer, B., \& Schmitt, J. H. M. M. 2004, A\&A, 418, 687
\bibitem{TDP04}
      Testa, P., Drake, J. J., \& Peres, G. 2004, ApJ, 617, 508
\end{thebibliography}
\end{document}